\documentstyle{article}
\title{\LARGE Multisoliton solutions of 3,4,6 waves problems connectet with semisimlealgebras of the second rank $A_3,B_2=C_2,G_2$}
\author{A.~N.~Leznov\thanks{ Universidad Autonoma del Estado de Morelos, CCICAp,Cuernavaca, Mexico}} \date{}

\begin{document}
\maketitle

\maketitle

\begin{abstract}

With each semisimple algebra it is possible to connect the system of interacting waves. The number of interacting fields coincides with the number of positive roots of corresponding semisimple algebra. Multisoliton solution of such kind problem are represented in explicit form for the algebras of second rank. 

\end{abstract}

\section{Introduction}

In the present paper we would like to constract multisoliton solutions for n- waves interacting system in explicit form, using technique of discrete transformation theory introduced in \cite{I}. The equation described n- waves interacting system are the following \cite{I}
\begin{equation}
[(dh),f_t]+[(ch),f_x]=[[(dh),f],[(ch),f]]  \label{MSC}
\end{equation}
where $f$ is algebra-valued unknown function (taking values in arbitrary semisimple algebra), $(x,t)$ independent arguments of the problem. In component form (\ref{MSC}) looks as
\begin{equation}
(c_R\frac{\partial}{\partial t}+d_R\frac{\partial}{\partial x})f_R=\sum_{P}(c_{R-P}d_P- d_{R-P}c_P)f_{R-P}f_P\label{CF}
\end{equation}

where by indexes $P,R$ are defined the set of all roots of semisimple algebra $c_P,d_P$ values of cartan elements $(ch),(dh)$ on these roots.
The case of $A_2$ algebra was considered in details in \cite{2} and multisoliton solutions in determinant form for this problem was present in \cite{TOK}. In \cite{TOK1} this results was rediscovered in the form which allow generalisation on the case of all semisimple algebras of the second rank (see sections below).  

\section{Solution of main equations in the case of of lower triangular algebra}

In this section we find solution of (\ref{CF}) under additional condition $f^+_R=0$ for all roots of semisimple algebra.
 
\subsubsection{Simple roots}

For each simple root equations (\ref{CF}) looks as
$$ 
(c_{\alpha}\frac{\partial}{\partial t}+d_{\alpha}\frac{\partial}{\partial x})f^-_{\alpha}=0
$$
with the obvious solution $f^-_{\alpha}=\Phi_{\alpha}(d_{\alpha}t-c_{\alpha}x)$. In the problem of the present paper for us it be essential parametrization of $f^-_{\alpha}$ in the form (of Lapalce transformation)
\begin{equation}
f^-_{\alpha}=\int d\lambda_{\alpha} s^{\alpha}(\lambda_{\alpha})e^{\lambda_{\alpha}(d_{\alpha}t-c_{\alpha}x)}\label{SR}
\end{equation}
Thus arbitarary simple root $\alpha$ of algebra is connected with arbitrary function of one variable
$s^{\alpha}(\lambda_{\alpha})$.

\subsubsection{Complicate root constructed from two simple ones}

The equation for such component is the following
$$
((c_{\alpha}+c_{\beta})\frac{\partial}{\partial t}+(d_{\alpha}+d_{\beta})\frac{\partial}{\partial x})f^-_{\alpha,\beta}=(c_{\alpha}d_{\beta}- d_{\alpha}c_{\beta})f^-_{\alpha}f^-_{\beta}
$$
Without any difficulties it is possible verified that solution of this equation is the following
\begin{equation}
f^-_{\alpha,\beta}=\int d\lambda_{\alpha}d\lambda_{\beta}
{s^{\alpha}(\lambda_{\alpha})s^{\beta}(\lambda_{\beta})\over \lambda_{\alpha}-\lambda_{\beta}}e^{t(d_{\alpha} \lambda_{\alpha}+d_{\beta} \lambda_{\beta})-x(c_{\alpha} \lambda_{\alpha}+c_{\beta} \lambda_{\beta})} \label{CR}   
\end{equation}
Plus solution of homogineous equation.

\subsubsection{Complicate root constructed from three simple ones}

It is possible two cases, when the fird root $\gamma$ not equal to one of the previous ones
$\alpha$ or $\beta$ or it coincedes with one of them. At first let us consider the first possibility. The equation (\ref{CF}) in this case looks as
$$
((c_{\alpha}+c_{\beta}+c_{\gamma})\frac{\partial}{\partial t}+(d_{\alpha}+d_{\beta}+d_{\gamma})\frac{\partial}{\partial x})f^-_{\alpha,\beta,\gamma}=
$$
\begin{equation}
(c_{\alpha}(d_{\beta}+d_{\gamma})- d_{\alpha}(c_{\beta}+c_{\gamma}))f^-_{\alpha}f^-_{\beta,\gamma}+((c_{\alpha}+c_{\beta})d_{\gamma})- (d_{\alpha}+d_{\beta})c_{\gamma})f^-_{\alpha,\beta}f^-_{\gamma}\label{DE}
\end{equation}
By direct check we find solution of the last equation in the form
$$
f^-_{\alpha,\beta,\gamma}=\int d\lambda_{\alpha}d\lambda_{\beta}d\lambda_{\gamma}
{s^{\alpha}(\lambda_{\alpha})s^{\beta}(\lambda_{\beta})s^{\gamma}(\lambda_{\gamma})\over (\lambda_{\alpha}-\lambda_{\beta})(\lambda_{\beta}-\lambda_{\gamma})}e^{t(d_{\alpha} \lambda_{\alpha}+d_{\beta} \lambda_{\beta}+d_{\gamma} \lambda_{\gamma})-x(c_{\alpha} \lambda_{\alpha}+c_{\beta} \lambda_{\beta}+c_{\gamma} \lambda_{\gamma})}    
$$
In the second case, when for instance $\beta=\gamma$ the defining equation (\ref{DE})does not 
contain one of the terms and in this case solution looks as 
$$
f^-_{\alpha,\beta,\beta}=\int d\lambda_{\alpha}d\lambda_{\beta_1}d\lambda_{\beta_2}
{s^{\alpha}(\lambda_{\alpha})s^{\beta}(\lambda_{\beta_1})s^{\beta}(\lambda_{\beta_2})\over (\lambda_{\alpha}-\lambda_{\beta_1})(\lambda_{\alpha}-\lambda_{\beta_2})}e^{t(d_{\alpha} \lambda_{\alpha}+d_{\beta} (\lambda_{\beta_1}+\lambda_{\beta_2}))-x(c_{\alpha} \lambda_{\alpha}+c_{\beta} (\lambda_{\beta_1}+\lambda_{\beta_2}))}    
$$

\section{The case of $A_2$ algebra}

This section is written only for convineous of the reader and contain necessary for what follows results of \cite{2},\cite{TOK1}. The results consernining $B_2$ and $G_2$ algebras will be presented below in the same form.

Algebra $A_2$  has the following Cartan matrix and basic commutation relations between two generators of the simple roots $X^{\pm}_{1,2}$ and its Cartan elements $h_{1,2}$
$$
k=\pmatrix{ 2 & -1 \cr
           -1 & 2 \cr},\quad \pmatrix{ [h_1,X^{\pm}_1]=\pm 2X^{\pm}_1 & [h_1,X^{\pm}_2]=\mp X^{\pm}_2 \cr
[h_2,X^{\pm}_1]=\mp X^{\pm}_2 & [h_2,X^{\pm}_2]=\pm 2X^{\pm}_2 \cr}
$$
Arbitrary element of the algebra may be represented as (up to Cartan elements which are absent in the problem under consideration)
$$
f=f^+_{1.1}X^+_{\alpha_1+\alpha_2}+f^+_{0.1}2X^+_{\alpha_2}+f^+_{1.0}X^+_
{\alpha_1}+f^-_{1.0}X^-_{\alpha_1}+f^-_{0.1}X^-_{\alpha_2}+f^-_{1.1}X^-_{\alpha_1+\alpha_2} 
$$
$\alpha_{1,2}$ are the indexes of si
mple roots. $X^+_{\alpha_1+\alpha_2}\equiv [X^+_2,X^+_1]$.

In these notations the system of equations
(\ref{CF}) looks as
$$
D_{1,0}f^+_{1.0}=f^+_{1.1}f^-_{0.1},\quad D_{1,0}f^-_{1.0}=f^-_{1.1}f^+_{0.1}
$$
\begin{equation}
D_{0,1}f^+_{0.1}=f^+_{1.1}f^-_{1.0},\quad D_{0,1}f^-_{0.1}=f^-_{1.1}f^+_{1.0}\label{DA}
\end{equation}
$$
D_{1,1}f^+_{1.1}=-f^+_{0.1}f^+_{1.0},\quad D_{1,1}f^-_{1.1}=-f^-_{0.1}f^-_{1.0}
$$
where operators of differenciation are the following ones $D_{i,j}={(ic_1+jc_2)\over \delta}\frac{\partial}{\partial t}+{(id_1+jd_2)\over \delta}\frac{\partial}{\partial x}$,
$\delta\equiv (c_1d_2-c_2d_1)$.

The discrete transformation of this system are the following ones \cite{2}

\subsubsection{ $T_3$}

The system (\ref{DA}) is invariant with respect to the following transformation $T_3$
$$
\tilde f^+_{1.1}={1\over f^-_{1.1}},\quad \tilde f^+_{1.0}=-{f^-_{0.1}\over f^-_{1.1}},\quad \tilde f^+_{0.1}={f^-_{1.0}\over f^-_{1.1}}
$$
\begin{equation}
\tilde f^-_{0.1}=-f^-_{1.1}D_{1.0}{f^-_{0,1}\over f^-_{1.1}},\quad \tilde f^-_{1.0}=f^-_{1.1}D_{0,1}{f^-_{1.0}\over f^-_{1.1}}\label{T3} 
\end{equation}
$$
{\tilde f^-_{1.1}\over f^-_{1.1}}=f^+_{1.1}f^-_{1.1}-D_{1,0}D_{0,1} \ln f^-_{1.1}
$$

\subsubsection{$T_2$}

The system (\ref{DA}) is invariant with respect to the following transformation $T_2$
$$
\tilde f^+_{0.1}={1\over f^-_{0.1}},\quad \tilde f^-_{1.0}=-{f^-_{1.1}\over f^-_{0.1}},\quad \tilde f^+_{12}={f^+_{1.0}\over f^-_{0.1}}
$$
\begin{equation}
\tilde f^+_{1.0}=-f^-_{0.1}D_{1,1}{f^+_{1.0}\over f^-_{0.1}},\quad \tilde f^-_{1.1}=-f^-_{0.1}D_{1,0}{f^-_{1.1}\over f^-_{0.1}}\label{T2} 
\end{equation}
$$
{\tilde f^-_{0.1}\over f^-_{0.1}}=f^+_{0.1}f^-_{0.1}+D_{1,0}D_{1,1} \ln f^-_{0.1}
$$

\subsubsection{$T_1$}

The system (\ref{DA}) is invariant with respect to the following transformation $T_1$
$$
\tilde f^+_{1.0}={1\over f^-_{1.0}},\quad \tilde f^-_{0.1}={f^-_{1.1}\over f^-_{1.0}},\quad \tilde f^+_{1.1}=-{f^+_{0.1}\over f^-_{1.0}}
$$
\begin{equation}
\tilde f^+_{0.1}=f^-_{0.1}D_{1,1}{f^+_{0.1}\over f^-_{1.0}},\quad \tilde f^-_{1.1}=f^-_{1.0}D_{0,1}{f^-_{1.1}\over f^-_{1.0}}\label{T1} 
\end{equation}
$$
{\tilde f^-_{1.0}\over f^-_{1.0}}=f^+_{1.0}f^-_{1.0}+D_{0,1}D_{1,1} \ln f^-_{1.0}
$$

\subsubsection{General properties of discrete transformations}

Three above transformations are invertable. This means $f$ may be expressed algebraicaly in terms of $\tilde f$. Exept of this $T_3=T_1T_2=T_2T_1$, what means that all discrete transformation are mutually commutative. This in its turn means that arbitrary discrete transformation may be represented in a form $T=T_1^{n_1}T_2^{n_2}$  \cite{2}.

\subsubsection{Result of consequent application of some number of discrete transformations}

In determinant form result was found in \cite{2}. In \cite{TOK1} it was rediscovered in the form which can be generalised to the case of arbitrary semi-simple algebra of second rank (see sections below).

Let us rewrite (\ref{SR}) and (\ref{CR}) in a form
$$
f^-_{1.0}=\int d\lambda p(\lambda)e^{\lambda(d_1t-c_1x)}\equiv \int d\lambda P(\lambda),\quad f^-_{0.1}=\int d\mu q(\mu)e^{\mu(d_2t-c_2x)}\equiv \int d\mu Q(\mu),
$$
$$
f^-_{1.1}=\int d\lambda\int d\mu {p(\lambda)q(\mu)\over \lambda-\mu}e^{\lambda(d_1t-c_1x)+\mu (d_2t-c_2x)}\equiv \int d\lambda\int d\mu {P(\lambda)Q(\mu)\over \lambda-\mu}
$$
and choose initial condition in a form
$$
f^+_{1.0}=f^+_{0.1}=f^+_{1.1}=0
$$
To solution of this kind application of each inverse transformation $T^{-1}_i$ is meanless wia zeros in denumerators.

Let us introduce determining function $U(n_1,n_2)$ 
\begin{equation}
U(n_1,n_2)=\int \prod_{i=1}^{n_1}P(\lambda_i) d\lambda_i \prod_{k=1}^{n_2} Q(\mu_k)d\mu_k {W_{n_1}^2(\lambda)W_{n_2}^2(\mu)\over \prod_{i,k} (\lambda_i-\mu_k)}\label{BU}
\end{equation}
where $W_n$ is Vandermond determinant constructed from $n_{1,2}$-variables $\lambda$ or $\mu$.

Then result of appliucation $n_1$ times discrete transformation $T_1$ and $n_2$ times discrete transformation $T_2$ looks as \cite{TOK1}   
$$
f^+_{1.0}={U(n_1-1,n_2)\over U(n_1,n_2)},\quad f^-_{1.0}={U(n_1+1,n_2)\over U(n_1,n_2)},
$$
\begin{equation}
f^+_{0.1}={U(n_1,n_2-1)\over U(n_1,n_2)},\quad f^-_{0.1}={U(n_1,n_2+1)\over U(n_1,n_2)},\label{GF}
\end{equation}
$$
f^+_{1.1}={U(n_1-1,n_2-1)\over U(n_1,n_2)},\quad f^-_{1.1}={U(n_1+1,n_2+1)\over U(n_1,n_2)}
$$
In the case of integrable systems connected with $A_1$ algebra Vandermond form for soliton solutions was introduced and used in \cite{4}.

Let us assume that initial functions has the form $P(\lambda)=\sum_{i=1}^{n_1} \delta(\lambda-l_k)e^{l_k(d_1t-c_1x)}$ and $Q(\mu)=\sum_{i=1}^{n_2} \delta(\mu-m_i)e^{m_i(d_2t-c_2x)}$, where $\delta(x)$ is the ussual Dirac $\delta$ function. In this case as it follows from (\ref{GF}) on the $n_1,n_2$ step of discrete transformation $f^-_{1.0}=f^-_{0.1}=f^-_{1.1}=0$ the chain is interrupted on the second end.
And from results of \cite{2} conditions of reality lead directly to soliton solutions.    

\section{The case of $B_2$ algebra}

Algebra $B_2$ (eqiuvalent to the algebra of the group of 5-dimentional rotations) has the following Cartan matrix and basics comutation relations
$$
k=\pmatrix{ 2 & -2 \cr
           -1 & 2 \cr},\quad \pmatrix{ [h_1,X^{\pm}_1]=\pm 2X^{\pm}_1 & [h_1,X^{\pm}_2]=\mp X^{\pm}_2 \cr
[h_2,X^{\pm}_1]=\mp 2 X^{\pm}_2 & [h_2,X^{\pm}_2]=\pm 2X^{\pm}_2 \cr}
$$
Arbitrary element of this algebra can be parametrised as (up to elements taking values in Cartan subalgebra)
$$
f=f^+_{1.2}X^+_{\alpha_1+2\alpha_2}+f^+_{1,1}X^+_{\alpha_1+\alpha_2}+f^+_{0.1}X^+_{\alpha_2}+f^+_{1.0}X^+_{\alpha_1}+f^-_{1.0}X^-_{\alpha_1}+f^-_{0.1}X^-_{\alpha_2}+f^-_{1.1}X^-_{\alpha_1+\alpha_2}+f^-_{1.2}X^-_{\alpha_1+2\alpha_2} 
$$
$\alpha_{1,2}$ are the indexes of simple roots. In these notations the system of equations
(\ref{CF}) looks as
$$
D_{1,0}f^+_{1.0}=2f^+_{1.1}f^-_{0.1},\quad D_{1,0}f^-_{1.0}=2f^-_{1.1}f^+_{0.1}
$$
$$
D_{0,1}f^+_{0.1}=f^+_{1.1}f^-_{1.0}+f^+_{1.2}f^-_{1.1},\quad D_{0,1}f^-_{0.1}=f^-_{1.1}f^+_{1.0}+f^-_{1.2}f^+_{1.1}
$$
\begin{equation}
{}\label{BW}
\end{equation}
$$
D_{1,1}f^+_{1.1}=-f^+_{0.1}f^+_{1.0}+f^+_{1.2}f^-_{0.1},\quad D_{1,1}f^-_{1.1}=-f^-_{0.1}f^-_{1.0}+f^-_{1.2}f^+_{0.1}
$$
$$
D_{1,2}f^+_{1,2}=-2f^+_{1.1}f^+_{0.1},\quad D_{1,2}f^-_{1.2}=-2f^-_{1.1}f^-_{0.1}
$$
where $D_{i,j}=({ic_1+jc_2\over \delta}\frac{\partial}{\partial t}+{id_1+jd_2\over \delta}\frac{\partial}{\partial x})$, $\delta\equiv (c_1d_2-c_2d_1)$ .
The above system of equations (\ref{BW}) is the wide system of interaction of eight fields $
f^{\pm}_{1.0},f^{\pm}_{0.1},f^{\pm}_{1.1},f^{\pm}_{1.2}$. This system is obviously allow reducing $(f^-)^H=f^+$ (under additional assumtion that all operators of differentiations are real). Exactly such kind solutions of this system is intersting for applications and will be the point of investigation in the present section.

The discrete transformation to the such kind of the systems in the case of arbitrary semisimple 
algebra was found in \cite{I}. The formulae below are partial case of application general construction for the case of $B_2$ algebra. They can be easely checked by direct not combersome calculation.

\subsection{Discrete transformation $T_M$}

The system (\ref{BW}) invariant with respect to following transformation $T_M$ (discrete transformation of the maximal root of $B_2$ algebra)
$$
\tilde f^+_{1.2}={1\over f^-_{1.2}},\quad \tilde f^+_{0.1}={f^-_{1.1}\over f^-_{1.2}},\quad \tilde f^+_{1.1}=-{f^-_{0.1}\over f^-_{1.2}},\quad \tilde f^+_{1.0}=f^+_{1.0}+{(f^-_{0.1})^2\over f^-_{1.2}},\quad \tilde f^-_{1.0}=f^-_{1.0}-{(f^-_{1.1})^2\over f^-_{1.2}}
$$
$$
\tilde f^-_2=-D_{1,0}f^-_2-f^+_{1.1}f^-_{1.2}+{1\over 2}f^-_{0.1}D_{1,0} \ln f^-_{1.2},\quad 
\tilde f^-_{1.1}=-D_{1,0}f^-_{1.1}+f^+_{0.1}f^-_{1.2}+{1\over 2}f^-_{1.1}D_{1,0} \ln f^-_{1.2}
$$
$$
{\tilde f^-_{1.2}\over f^-_{1.2}}={1\over 4}D_{1,0}D_{1,0} \ln f^-_{1.2}+{f^-_{1.1}D_{1,0}f^-_{0.1}-f^-_{0.1}D_{1,0}f^-_{1.1}\over 2f^-_{1.2}}+f^+_{1.2}f^-_{1.2}+f^+_{1.1}f^-_{1.1}+f^+_{0.1}f^-_{0.1}
$$
By direct not combersome calculations it is possible verified that $\tilde f$ satisfy (\ref{BW})
if $f$ is solution of this system.

\subsection{Discrete transformation $T_{1.0}$}

$T_{1.0}$-discrete transformation of the first ($\alpha_1$) root of $B_2$ algebra
$$
\tilde f^+_{1.0}={1\over f^-_{1.0}},\quad \tilde f^-_{0.1}={f^-_{1.1}\over f^-_{1.0}},\quad \tilde f^+_{1.1}=-{f^+_{0.1}\over f^-_{1.0}},\quad \tilde f^+_{1.2}=f^+_{1.2}+{(f^+_{0.1})^2\over f^-_1},\quad \tilde f^-_{1.2}=f^-_{1.2}-{(f^-_{1.1})^2\over f^-_{1.0}}
$$
\begin{equation}
\tilde f^+_{0.1}=D_{1,2}f^+_{0.1}-f^+_{1.1}f^-_{1.0}-{1\over 2}f^+_{0.1}D_{1,2} \ln f^-_{1.0},\quad \tilde f^-_{1.1}=D_{1,2}f^-_{1,1}+f^-_{0.1}f^-_{1.0}-{1\over 2}f^-_{1.1}D_{1,2} \ln f^-_{1.0}\label{R1}
\end{equation}
$$
{\tilde f^-_{1.0}\over f^-_{1.0}}={1\over 4}D_{1,2}D_{1.2} \ln f^-_{1.0}+{f^-_{1.1}D_{1,2}f^+_{0.1}-f^+_{0.1}D_{1,2}f^-_{1.1}\over 2f^-_{1.0}}+f^+_{1.0}f^-_{1.0}-f^+_{1.1}f^-_{1.1}+f^+_{0.1}f^-_{0.1}
$$

The most important propertie of this transformation consists in the fact that if we begin from the initial solution $f^+_{1.0}=f^+_{0.1}=f^+_{1.1}=f^+_{1.2}=0$ the last three function contains there zero values after discrete transformation $T_{1.0}$ as it follows from (\ref{R1}). 

\subsection{Inverse discrete transformation $T^{-1}_{1.0}$}

To construct inverse $T^{-1}_{1.0}$ it is necessary formulae (\ref{R1}) resolve with respect to untilded functions. Result is the the following one 
$$
f^-_{1.0}={1\over \tilde f^+_{1.0}},\quad f^+_{0.1}=-{\tilde f^+_{1.1}\over \tilde f^+_{1.0}},\quad  f^-_{1.1}={\tilde f^-_{0.1}\over \tilde f^+_{1.0}},\quad f^+_{1.2}=\tilde f^+_{1.2}-{(\tilde f^+_{1.1})^2\over \tilde f^+_{1.0}},\quad  f^-_{1.2}=\tilde f^-_{1.2}+{(\tilde f^-_{0.1})^2\over \tilde f^+_{1.0}}
$$
\begin{equation}
f^-_{0.1}=-D_{1,2}\tilde f^-_{0.1}-\tilde f^-_{1.1}\tilde f^+_{1.0}-{1\over 2}\tilde f^-_{0.1}D_{1,2} \ln \tilde f^+_{1.0},\quad f^+_{1.1}=-D_{1,2}\tilde f^+_{1.1}+\tilde f^+_{0.1}\tilde f^+_{1.0}-{1\over 2}\tilde f^+_{1.1}D_{1,2} \ln \tilde f^+_{1.0}\label{R2}
\end{equation}
$$
{f^+_{1.0}\over \tilde f^+_{1.0}}={1\over 4}D_{1,2}D_{1,2} \ln \tilde f^+_{1.0}-{\tilde f^+_{1.1}D_{1,2}\tilde f^-_{0.1}-\tilde f^-_{0.1}D_{1,2}\tilde f^+_{1.1}\over 2\tilde f^+_{1.0}}+\tilde f^+_{1.0}\tilde f^-_{1.0}+\tilde f^+_{1.1}\tilde f^-_{1.1}+\tilde f^+_{0.1}\tilde f^-_{0.1}
$$

\subsection{Discrete transformation $T_{2\alpha_2}$}

In the case of $B_2$ algebra it is not possible to coinside with the help of the Weil group
the maximal root with $X^{\pm}_{\alpha_2}$. But having in mind $T_M$ and $T_1$ it is possible to construct dicrete transformation $T_{2\alpha_2}\equiv T_{\alpha_1+2\alpha_2}T^{-1}_{\alpha_1}=
T_M T^{-1}_1$ which we will call discrete transformation of the second root and which looks as
$$
\tilde f^-_{1.0}={1\over f^+_{1.0}+{(f^-_{0.1})^2\over f^-_{1.2}}},\quad \tilde f^-_{1.1}={-D_{1.0}f^-_{0.1}-f^+_{1.1}f^-_{1.2}+{1\over 2}f^-_{0.1}D_{1.0} \ln f^-_{1.2}\over f^+_{1.0}+{(f^-_{0.1})^2\over f^-_{1.2}}},\quad \tilde f^+_{0.1}={f^-_{0.1}\over f^+_{1.0}f^-_{1.2}+(f^-_{0.1})^2}
$$
$$
\tilde f^+_{1.2}={f^+_{1.0}\over f^+_{1.0}f^-_{1.2}+(f^-_{0.1})^2},\quad \tilde f^+_{1.1}={f^+_{1.0}D_{1.2}f^-_{0.1}-f^-_{1.1}(f^+_{1.0})^2-{1\over 2}f^-_{0.1}D_{1.2}f^+_{1.0} \over f^+_{1.0}+{(f^-_{0.1})^2\over f^-_{1.2}}}
$$
From the last formulae it follows that $T_{2\alpha_2}$ conserve zero values for wave functions
$f^+_{1.0}=f^+_{1.1}=f^+_{1.2}=0$ under initial conditions under consideration. By this reason we present further formulae after substitution in them these values. Because the general formulae are suffitiently combersom.
$$
{\tilde f^-_{0.1}\over f^-_{0.1}}=D_{1,0}^2 \ln f^-_{0.1}+f^-_{0.1}f^+_{0.1},\quad \tilde f^+_{0.1}={1\over f^-_{0.1}}
$$
This is the equations of the ussual Toda lattice solution of which are well known. And at last after some number of manipulations we obtain
$$
\tilde f^-_{1.2}={1\over 4}D_{1,0}^2 f^-_{1.2}+{f^-_{1.1}D_{1,0} f^-_{0.1}-f^-_{0.1}D_{1,0} f^-_{1.1}\over 2}+
$$
$$
{D_{1,0}f^-_{0.1}D_{1,0} f^-_{0.1}\over (f^-_{0.1})^2}f^-_{1.2}-{D_{1,0}f^-_{0.1}D_{1.0} f^-_{1.2}\over f^-_{0.1}}+f^+_{0.1}f^-_{0.1}f^-_{1.2}
$$

\subsection{Linear integrable chain of the second simple root in the field of Toda lattice}

Let us consider consequent action of discrete transformation of the previous subsection on the initial solution $f^+_{1.0}=f^+_{0.1}=f^+_{1.1}=f^+_{1.2}=0$. As it fololows from explicit formulae for discrete transformation zero values for $f^+_{1.0}=f^+_{1.1}=f^+_{1.2}=0$ are conserved. Equations for $f^-_{0.1},f^+_{0.1}$ functions are typical one dimentional Toda chain with well known solution. Result of the application n-times of Toda discrete transformation to initial functon $f^-_{0.1}\equiv r$ is the following
\begin{equation}
f^+_{0.1}={Det_{n-1}\over Det_n} ,\quad f^-_{0.1}={Det_{n+1}\over Det_n}\label{TO}
\end{equation}
where $Det_n$ are the the main minores from left upper corner of the matrix
\begin{equation}
\Delta=\pmatrix{ r & r_1 & r_{11} & .....\cr
							 r_1 & r_{11} & r_{111} & .....\cr
							r_{11} & r_{111} & r_{1111} & .....\cr
							...... & ....... & .........& .....\cr}\label{DM}
\end{equation}
where $r_1\equiv D_{1,0}r$ and so on.  
In these notations equations of Toda latice looks as
\begin{equation}
 D^2_{1,0}(\ln D_n)={Det_{n-1}Det_{n+1}\over Det^2_n} \label{TLS}
\end{equation}
Initial function of the second simple root $r=f^-_{0.1}=\int d\mu q(\mu)e^{\mu(d_2t-c_2x)}\equiv \int d\mu Q$. In $Q$ all dependence on arguments $x,t$ is included. Obviously
$D_{0,1}Q=0$. In these notations $Det_n$ may be presented in the form of multidimensional                                                    integrale
$$
Det_n={1\over n!}\int \prod_{i=1}^n d\mu_i Q_{\mu_i} W^2_n(\mu_1,..,\mu_n)
$$
where $W^2_n(\mu_1,..,\mu_n)$ is squere of Vandermond determinant.

\subsubsection{Resolving $T_{2\alpha_2}$ discrete transformation chain} 

The $T_{2\alpha_2}$ discrete transformation may be presented as linear chain of equations for only two unknown functions $f^-_{1.1},f^-_{1.2}$. Paramerizing them in the form
$(f^-_{1.1})_n={A^n\over D_n^2},\quad (f^-_{1.2})_n={B^n\over D_n^2}$ we pass to the following chain of equations 
\begin{equation}
A^{n+1}={{1\over 2}D_{n+1}B^n_1-B^n(D_{n+1})_1\over D_n}\label{CA}
\end{equation}
$$  
{B^{n+1}\over D^2_{n+1}}={1\over D_n^2}[{1\over 4} B^n_{11}-{(D_{n+1})_1\over D_{n+1}}B^n_1+
({(D_{n+1})_1\over D_{n+1}})^2B^n]+
$$
\begin{equation}
{}\label{CB}
\end{equation}
$$
{A^n (D_{n+1})_1-D_{n+1}A^n_1\over 2D_n^3}+{D_{n+1}(A^n (D_n)_1+B^n D_{n-1})\over 2D_n^4}
$$
The last unknown function in these notations looks as $f^-_{1.0}={B^{n-1}\over D^2_n}$.

Let us resolve (\ref{CB}) by induction, assuming (see appendix), that
\begin{equation}
A^n={1\over n!(n+1)!} \int P(\lambda)d\lambda \prod_{i=1}^{(2n+1)}Q(\mu_i) d\mu_i {W^2_n(\mu_1,..\mu_n)W^2_{n+1}(\mu_{n+1},..\mu_{2n+1})\over \prod_{i=1}^{(2n+1)}(\lambda-\mu_i)}\label{A}
\end{equation}
\begin{equation}
B^n={1\over (n+1)!(n+1)!}\int P(\lambda)d\lambda \prod_{i=1}^{(2n+2)}Q(\mu_i) d\mu_i {W^2_{n+1}(\mu_1,..\mu_{n+1})W^2_{n+1}(\mu_{n+2},..\mu_{2n+2})\over \prod_{i=1}^{(2n+2)}(\lambda-\mu_i)}\label{B}
\end{equation}
From this moment all factorial factors are included in definition of Vandermond determinant.
Substituting assumed form of solution into (\ref{CA}) we obtain integrant function in a form
$$
{W^2_{n+1}(\mu_1,..\mu_{n+1})W^2_{n+1}(k_1,..k_{n+1})\over \prod_{i=1}^{(n+1)}(\lambda-\mu_i)(\lambda-k_i)}
({1\over 2}\sum_{i=1}^{n+1}(\mu_i+k_i)-\sum_{i=1}^{n+1}d_i)W^2_{n+1}(d_1,....d_{n+1})
$$
The sum with factor ${1\over 2}$ arises after differentiation $B^n$ on $D_{1,0}$. The second sum arises after differentiation of $D_{n+1}$ on the same operator. But domain of integration is symmetrical with respect $(2n+2)$ parameters $(m,k)$ and $(n+1)$ parameters $d$. Thus it is possible rewrite the last expression in a form
\begin{equation}
{W^2_{n+1}(\mu_1,..\mu_{n+1})W^2_{n+1}(k_1,..k_{n+1})\over \prod_{i=1}^{(n+1)}(\lambda-\mu_i)(\lambda-k_i)}(\sum_1^{n+1}(\mu_i-d_i)W^2_{n+1}(d_1,....d_{n+1})\label{MID}
\end{equation}
Now let us compare the last expression with solution in the case of $A_2$ arising after application n times discrete transformation $T_2$. For interesting for us functions in connection with (\ref{GF}) we have
$$
f^-_{1.0}={V(1,n+1)\over V(0,n+1)}\quad, f^-_{1.1}={V(1,n+2)\over V(0,n+1)},\quad 
f^+_{0.1}={V(0,n)\over V(0,n+1)}
$$
We rewrite now equation for $f^-_{1.0}$ (\ref{DA}) conserving only integrant function in both
sides
$$
{W^2_{n+1}(\mu_1,..\mu_{n+1})\over \prod_{i=1}^{(n+1)}(\lambda-\mu_i)}
(\sum_{i=1}^{n+1}\mu_i-\sum_{i=1}^{n+1}d_i)W^2_{n+1}(d_1,....d_{n+1})=
$$
\begin{equation}
{W^2_{n+2}(\mu_1,..\mu_{n+2})\over \prod_{i=1}^{(n+2)}(\lambda-\mu_i)}
W^2_n(d_1,....d_n)\label{GRA})
\end{equation}
where $\mu_{n+2}=d_{n+1}$ (to have the same integrable indexes in both sides of the last equality).
The last equality after substitution it into (\ref{MID}) finish the proof of reccurent relation for $A^{n+1}$ term of the chain above. 

Now we pass to calculation of $B^{n+1}$ defined by (\ref{CB}). The terms in the second row of this equality may be rewrriten in a equivalent form
$$
{A^n (D_{n+1}D_n)_1-D_{n+1}D_n A^n_1+B^n D_{n+1}D_{n-1}\over 2D_n^4}
$$
Substituting into the last expression  assuming by reduction form of $A^n$ and using once more
(\ref{GRA}) we represent it as
$$
-{\int P(\lambda)d\lambda \prod_{i=1}^{(n+2)}Q(\mu_i)d\mu_i \prod_{i=1}^{n}Q(\nu_i) d\nu_i {W^2_n(\nu_1,..\nu_n)W^2_{n+2}(\mu_1,..\mu_{n+2})\over \prod_{i=1}^{(n+2)}(\lambda-\mu_i)\prod_{k=1}^n(\lambda-\nu_k)}\over 2D_n^2}
$$
The summation the terms of the first row of (\ref{CB}) leads to the following result in numenator
$$
{W^2_{n+1}(\mu_1,..\mu_{n+1})\over \prod_{i=1}^{(n+1)}(\lambda-\mu_i)}
{W^2_{n+1}(\nu_1,....\nu_{n+1})\over \prod_{i=1}^{(n+1)}(\lambda-\mu_i)}
$$
\begin{equation}
(\sum^{n+1}_1({\mu_i+\nu_i\over 2}-\sigma_i)\sum^{n+1}_1({\mu_i+\nu_i\over 2}-\delta_i)W^2_{n+1}(\sigma_1,..\sigma_{n+1})W^2_{n+1}(\delta_1,..\delta_{n+1})\label{XO}
\end{equation}
In denumerator we have $D^2_n D^2_{n+1}$. The origin of the terms in (\ref{XO}) is as follows.
Each term contain linear $B^n$ and its derivatives. We parametrised this function by $\nu,\mu$ parameters. Each term contains second degree of $D_{n+1}$ or its derivatives. We parametrised
them by independent parameters $\sigma,\delta)$. The first derivatives of $B^n$ function leads to multiplicator $\sum^{n+1}_1(\mu_i+\nu_i)$. Keeping in mind the factor ${1\over 4}$ at the first term in (\ref{CB}) we explain corresponding term in (\ref{XO}). Quadratical in derivades
on $(D_{n+1})_1$ terms leads obviously to product of sums of $\sigma$ and $\delta$ parameters and so on. 

Now stratedgy of the further calculatios will be the following. We multiply both sides of (\ref{CB}) on $D_n^2$ and use once more (\ref{GRA}) in the back direction we come to the following expression in the left side:
$$
{W^2_{n+2}(\mu_1,..\mu_{n+2})\over \prod_{i=1}^{(n+2)}(\lambda-\mu_i)}W^2_n(\sigma_1,....\sigma_n){W^2_{n+2}(\nu_1,..\nu_{n+2})\over \prod_{i=1}^{(n+2)}(\lambda-\nu_i)}W^2_n(\delta_1,....\delta_n)
$$
where determinant of Vandermond parametrized by $\mu,\nu$ arised from the assumed form of $B^{n+1}$ (\ref{B}), parametrised by $\delta,\sigma$ ones  from $D_n^2$. Now let us use
(\ref{GRA}) twise. We obtain
$$
{W^2_{n+1}(\mu_1,..\mu_{n+1})\over \prod_{i=1}^{(n+1)}(\lambda-\mu_i)}
(\sum_{i=1}^{n+1}\mu_i-\sum_{i=1}^{n+1}\delta_i)W^2_{n+1}(\delta_1,....\delta_{n+1})\times
$$
$$
{W^2_{n+1}(\nu_1,..\nu_{n+1})\over \prod_{i=1}^{(n+1)}(\lambda-\nu_i)}
(\sum_{i=1}^{n+1}\nu_i-\sum_{i=1}^{n+1}\sigma_i)W^2_{n+1}(\sigma_1,....\sigma_{n+1})
$$
The structure of last expression exactly the same as in (\ref{XO})- the combination terms in the right side of (\ref{B}). Different only multiplicators. Differenece of them is the following
$(\mu\equiv \sum_{i=1}^{n+1}\mu_i$ and so on)
$$
({\mu+\nu\over 2}-\sigma)({\mu+\nu\over 2}-\delta)-(\mu-\delta)(\nu-\sigma)=({\mu-\nu\over 2})^2
$$

In the last expression dependence on parameteres $\delta,\sigma$ is factorised and leads to $D^2_{n+1}$ which is canceled with the same term in denumeratoor. The remaining relation, which have been proved is the following
$$
{W^2_{n+1}(\nu_1,..\nu_{n+1})\over \prod_{i=1}^{(n+1)}(\lambda-\nu_i)}
({\sum_{i=1}^{n+1}\nu_i-\sum_{i=1}^{n+1}\mu_i\over 2})^2{W^2_{n+1}(\mu_1,....\mu_{n+1})\over \prod_{i=1}^{(n+1)}(\lambda-\mu_i)}-
$$
$$
-{1\over 2}{W^2_n(\nu_1,..\nu_n)W^2_{n+2}(\mu_1,..\mu_{n+2})\over \prod_{i=1}^{(n+2)}(\lambda-\mu_i)\prod_{k=1}^n(\lambda-\nu_k)}=0
$$
But the last equality exactly coincedes with equations of the Toda lattice (\ref{TLS}) 
Thus the solution of the linear chain (\ref{CA}) and (\ref{CB})is proved by induction.
Now to this solution it is necessary applicate $n_1$ times discrete transformation of the first
simple root $T_1$. We will not do corresponding calculations but in the next subsection will consider discrete transformation chain of the first simple root. From this consideration general form of solution will be obvious.

\subsubsection{Resolving $T_{1.0}$ discrete transformation chain}
 
As it was mentioned after explicit formulae of discrete transformation (\ref{R1})
after action of this transformation on the initial functions $f^+_{1.0}=f^+_{0.1}=f^+_{1.1}=f^+_{1.2}=0$  the last three functions conserved there zero values. The remaing equations describing arising chain looks as
$$
f^+_{1.0}={1\over \tilde f^-_{1.0}},\quad  \tilde f^-_{0.1}={\tilde f^-_{1.1}\over  f^-_{1.0}},\quad \tilde f^-_{1.2}=f^-_{1.2}-{(f^-_{1.1})^2\over \tilde f^-_{1.0}}
$$
\begin{equation}
\tilde f^-_{1.1}=D_{1,2}f^-_{1.1}+f^-_{0.1}f^-_{1.0}-{1\over 2}f^-_{1.1}D_{1,2} \ln  f^-_{1.0}\label{R3}
\end{equation}
$$
{\tilde f^-_{1.0}\over f^-_{1.0}}={1\over 4}D_{1,2}D_{1,2} \ln f^-_{1.0}+f^+_{1.0}f^-_{1.0}
$$
Now it is necessary to take into acount that $D_{1,0}f^-_{1.0}=0$ and in connection of equations of motion (\ref{BW}) $D_{1,1}f^-_{1.0}=-f^-_{0.1}f^-_{1.0}$ (the last term $f^-_{1.2}f^+_{0.1}=0$) we rewrite the last two equations from (\ref{R3}) in a equivalent form
$$
\tilde f^-_{1.1}=f^-_{1.0}D_{0,1}{f^-_{1.1}\over f^-_{1.0}}
$$
$$
{\tilde f^-_{1.0}\over f^-_{1.0}}=D_{0,1}^2 \ln f^-_{1.0}+f^+_{1.0}f^-_{1.0},\quad f^+_{1.0}={1\over \tilde f^-_{1.0}}
$$
The system of equation above for $f^+_{1.0},f^-_{1.0},f^-_{1.1}$ functions exactly coinsides with discrete transformation $T_1$ (\ref{T1}) for the first simple root of $A_2$ algebra solition of which was found above in the third section. The last function $f^-_{1.2}$ is defined algebraicaly via $f^-_{1.1},f^-_{1.0}$ from the first row of (\ref{R3}). 
By reduction using only first equality of Yacobi \cite{2} we obtain
\begin{equation}
f^-_{1.1}={\tilde Det_{n+1}\over Det_n},\quad f^-_{1.2}={\tilde {\tilde D}et_{n+1}\over Det_n}\label{FINAL}
\end{equation}
where $Det_n$ are the main minores of the matrix
\begin{equation}
\pmatrix{ f^-_{1.0} & D_{0,1}f^-_{1.0} & D_{0,1}^2f^-_{1.0} & .....\cr
	D_{0,1} f^-_{1.0} & D_{0,1}^2 f^-_{1.0} & D_{0,1}^3 f^-_{1.0} & .....\cr
	D_{0,1}^2 f^-_{1.0} & D_{0,1}^3 f^-_{1.0} & D_{0,1}^4 f^-_{1.0} & .....\cr
						...... & ....... & .........& .....\cr}\label{YOB}
\end{equation}
$\tilde D_n$ determinant of matrix (\ref{YOB}) in which last column is constructed from $f^-_{1.1}$ function and its consequent derivatives. And at last $\tilde {\tilde D}_n$ in matrix
(\ref{YOB}) the last column and row are exchanged on consequent deritivies of $f^-_{1.1}$ function exept of diagonal term which is occupied by $f^-_{1.2}$ function. One example clarify the situation
$$
\tilde {\tilde D_3}=Det_3\pmatrix{ f^-_{1.0} & D_{0,1}f^-_{1.0} & f^-_{1.1} \cr
	D_{0,1}f^-_{1.0} & D_{0,1}^2f^-_{1.0} & D_{0,1}f^-_{1.1}\cr
	f^-_{1.1} & D_{0,1}f^-_{1.1} & f^-_{1.2} \cr}
$$					
Integrales in the (\ref{FINAL}) may be calculated directly by methods of \cite{TOK1}. We present calculations for more complicate case of $\tilde {\tilde D}et_{n+1}$. We paramerise first $n$
elements of the first column by $\lambda_1$ (parameter of integration) and the last term $f^-_{1.1}$ by two parameters $\lambda_1,\mu_1$. Elements of second column are parametrised by
$\lambda_2,\mu_1$ and so on up to $n-th$ column $\lambda_n,\mu_1$. Elements of $(n+1)$-th column
n first one by $\lambda_{n+1},\mu_2$ and the last one  $f^-_{1.2}$ by three parameters $\lambda_{n+1},\mu_1,\mu_2$. As a result under the sign of multidimensional ($(n+3)$) integral
we obtain determinant, which may be easely calculated by the same method as calculated Vandermond determinant. Result is as follows (we present integrand function)
$$
\tilde {\tilde D_{n+1}}={W^2_{n+1}(\lambda)\over \prod_{i=1}^{(n+1)}(\lambda_i-\mu_1)(\lambda_i-\mu_2)},\quad \tilde D_{n+1}={W^2_{n+1}(\lambda)\over \prod_{i=1}^{(n+1)}(\lambda_i-\mu)}
$$
Keeping in mind that transformation $T_{2\alpha_2}$ change only $n_2$ and transformation $T_1$
changes only $n_1$ we come to the solution of the next subsection

\subsection{General formulae for solution}

The result of calculation of discretre transformation $T_{2\alpha_2}^{n_2}T_1^{n_1}$ may be expessed in terms of the basis function
$$
V(n_1;n_2,n_3)={1\over n_1!n_2!n_3!} \int \prod_{k=1}^{n_1}P(\lambda_k) d\lambda_k \prod_{i=1}^{n_2)}Q(\mu_i) d\mu_i\prod_{j=1}^{n_3)}Q(\nu_j) d\nu_j \times
$$
$$
{W^2(\lambda)_{n_1}W^2(\mu)_{n_2}W^2(\nu)_{n_3}\over \prod_{k=1}^{n_1}\prod_{i=1}^{n_2}\prod_{j=1}^{n_3}(\lambda_k-\mu_i)(\lambda_k-\nu_j)}
$$
In this notations solution of 4-wave $B_2$ problem looks similar as it was found before for the case of $A_2$ algebra
$$
f^+_{1.0}={V(n_1-1;n_2,n_2)\over V(n_1;n_2,n_2)},\quad f^+_{0.1}={V(n_1;n_2,n_2-1)\over V(n_1;n_2,n_2)}
$$
$$
f^+_{1.1}={V(n_1-1;n_2,n_2-1)\over V(n_1;n_2,n_2)},\quad f^+_{1.2}={V(n_1-1;n_2-1,n_2-1)\over V(n_1;n_2,n_2)}
$$
\begin{equation}
f^-_{1.0}={V(n_1+1;n_2,n_2)\over V(n_1;n_2,n_2)},\quad f^-_{0.1}={V(n_1;n_2+1,n_2)\over V(n_1;n_2,n_2)}\label{V}
\end{equation}
$$
f^-_{1.1}={V(n_1+1;n_2+1,n_2)\over V(n_1;n_2,n_2)},\quad f^-_{1.2}={V(n_1+1;n_2+1,n_2+1)\over V(n_1;n_2,n_2)}
$$
In the case $n_1=0$ this exactly solution of chain of the second root; in the case $n_2=0$ this is solution of the chain of the first root.

\section{The case of $G_2$ algebra}

Algebra $G_2$  has the following Cartan matrix and basic commutation relations between two generators of the simple roots $X^{\pm}_{1,2}$ and its Cartan elements $h_{1,2}$
$$
k=\pmatrix{ 2 & -3 \cr
           -1 & 2 \cr},\quad \pmatrix{ [h_1,X^{\pm}_1]=\pm 2X^{\pm}_1 & [h_1,X^{\pm}_2]=\mp X^{\pm}_2 \cr
[h_2,X^{\pm}_1]=\mp 3X^{\pm}_2 & [h_2,X^{\pm}_2]=\pm 2X^{\pm}_2 \cr}
$$
System of its positive roots contain 6 elements
$$
X^+_1=X^+_{\alpha_1},X^+_2=X^+_{\alpha_2},X^+_{1;2}=[X^+_2,X^+_1]=X^+_{\alpha_1+\alpha_2},
X^+_{1;22}={1\over 2}[X^+_2[X^+_2,X^+_1]]=X^+_{\alpha_1+2\alpha_2},
$$
$$
X^+_{1;222}=={1\over 6}[X^+_2[X^+_2[X^+_2,X^+_1]]]=X^+_{\alpha_1+3\alpha_2},X^+_{11;222}={1\over 6}[X^+_1[X^+_2[X^+_2[X^+_2,X^+_1]]]]=X^+_{2\alpha_1+3\alpha_2}
$$
The coefficients of decomposition by these roots will be denoted as $f_{n;m}$ correspondingly.
System of equations for 12 functions $f^{\pm}_{2.3},f^{\pm}_{1.3},f^{\pm}_{1.2},f^{\pm}_{1.1},f^{\pm}_{1.0},f^{\pm}_{0.1}$ looks as
$$
D_{2,3}f^+_{2.3}=3f^+_{1.0}f^+_{1.3}-3f^+_{1.1}f^+_{1.2},\quad D_{2,3}f^-_{2.3}=3f^-_{1.0}f^-_{1.3}-3f^-_{1.1}f^-_{1.2}
$$
$$
D_{1,3}f^+_{1.3}=-3f^+_{2.3}f^-_{1.0}-3f^+_{0.1}f^+_{1.2},\quad D_{1,3}f^-_{1.3}=-3f^+_{1.0}f^-_{2.3}-3f^-_{0.1}f^-_{1.2}
$$
$$
D_{1,2}f^+_{1.2}=f^+_{2.3}f^-_{1.1}+f^+_{1.3}f^-_{0.1}-2f^+_{0.1}f^+_{1.1},\quad D_{1,2}f^-_{1.2}=f^-_{2.3}f^+_{1.1}+f^-_{1.3}f^+_{0.1}-2f^-_{0.1}f^-_{1.1}
$$
\begin{equation}
{} \label{SG}
\end{equation}
$$
D_{1,1}f^+_{1.1}=f^+_{2.3}f^-_{1.2}+2f^+_{1.2}f^-_{0.1}-f^+_{0.1}f^+_{1.0},\quad D_{1,1}f^-_{1.1}=f^-_{2.3}f^+_{1.2}+2f^-_{1.2}f^+_{0.1}-f^-_{0.1}f^-_{1.0} 
$$
$$
D_{1,0}f^+_{1.0}=-3f^+_{2.3}f^-_{1.3}+3f^+_{1.1}f^-_{0.1},\quad D_{1,0}f^-_{1.0}=-3f^-_{2.3}f^+_{1.3}+3f^-_{1.1}f^+_{0.1}
$$
$$
D_{0,1}f^+_{0.1}=f^+_{1.3}f^-_{1.2}+2f^+_{1.2}f^-_{1.1}+f^+_{1.1}f^-_{1.0},\quad D_{0,1}f^-_{0.1}=f^-_{1.3}f^+_{1.2}+2f^-_{1.2}f^+_{1.1}+f^-_{1.1}f^+_{1.0}
$$

\subsection{Initial conditions for $G_2$}

Let us rewrite (\ref{SR}) and (\ref{CR}) in a form under condition that all $f^+=0$
$$
f^-_{1.0}=\int d\lambda p(\lambda)e^{\lambda(c_1x-d_1t)},\quad f^-_{0.1}=\int d\mu q(\mu)e^{\mu(c_2x-d_2t)},
$$
$$
f^-_{1.1}=\int d\lambda\int d\mu {p(\lambda)q(\mu)\over \lambda-\mu}e^{\lambda(c_1x-d_1t)+\mu(c_2x-d_2t)}
$$
$$
f^-_{1.2}=-\int d\lambda\int d\mu_1 d\mu_2 {p(\lambda)q(\mu_1)q(\mu_2)\over (\lambda-\mu_1)(\lambda-\mu_2)}e^{\lambda(c_1x-d_1t)+(\mu_1+\mu_2)(c_2x-d_2t)}
$$
$$
f^-_{1.3}=\int d\lambda\int d\mu_1 d\mu_2 d\mu_3 {p(\lambda)q(\mu_1)q(\mu_2)q(\mu_3)\over (\lambda-\mu_1)(\lambda-\mu_2)(\lambda-\mu_3)}e^{\lambda(c_1x-d_1t)+(\mu_1+\mu_2+\mu_3)(c_2x-d_2t)}
$$
$$
f^-_{2.3}={1\over 2}\int d\lambda_1d\lambda_2 (\lambda_1-\lambda_2)^2\int d\mu_1 d\mu_2 d\mu_3 {p(\lambda_1)p(\lambda_2)q(\mu_1)q(\mu_2)q(\mu_3)\over \prod_{i=1,a=1}^{i=2,a=3}(\lambda_i-\mu_a)}e^R
$$
$$
R=(\lambda_1+\lambda_2)(c_1x-d_1t)+(\mu_1+\mu_2+\mu_3)(c_2x-d_2t)
$$

\subsection{Discrete transformation of the first simple root $T_1$}
 
We remind to the reader that first seven equation of discrete transformation below is a direct consequence of results of the paper \cite{I}. Of course it is possible consider them as a happy guess. The most simple way of obtaining  last five equation consists in direct substitution the 7 previous ones in the system (\ref{SG}). 
$$
\tilde f^+_{1.0}={1\over f^-_{1.0}},\quad \tilde f^-_{1.3}=-{f^-_{2.3}\over f^-_{1.0}},\quad 
\tilde f^+_{1.1}=-{f^+_{0.1}\over f^-_{1.0}},\quad \tilde f^-_{0.1}={f^-_{1.1}\over f^-
_{1.0}},\quad \tilde f^+_{2.3}={f^+_{1.3}\over f^-_{1.0}}
$$
$$
\tilde f^+_{1.2}=f^+_{1.2}+{f^-_{1.1}f^+_{1.3}+(f^+_{0.1})^2\over f^-_{1.0}},\quad  
\tilde f^-_{1.2}=f^-_{1.2}-{f^-_{2.3}f^+_{0.1}+(f^-_{1.1})^2\over f^-_{1.0}}
$$
The expressions above now it is possible to substitute into (\ref{SG}) and obtain all other equations of discrete transformation. For instant rewriting equation for $f^-_{1.3}$ function for transformed variables
$$
D_{1,3}\tilde f^-_{1.3}=-3\tilde f^+_{1.0}\tilde f^-_{2.3}-3\tilde f^-_{0.1}\tilde f^-_{1.2}
$$
and substituting in in it transformed functions from above equations after not combersome calculations we find $\tilde f^-_{2.3}$ and so on with the result
$$
\tilde f^-_{2.3}={f^-_{1.0}D_{1,2}f^-_{2.3}-{1\over 2}f^-_{2.3}D_{1,2}f^-_{1.0}\over f^-_{1.0}}-
f^-_{1.0}f^-_{1.3}+{-(f^-_{2.3})^2f^+_{1.3}+2(f^-_{1.1})^3+3f^-_{2.3}
f^-_{1.1}f^+_{0.1}\over 2f^-_{1.0}}
$$
$$
\tilde f^+_{0.1}={f^-_{1.0}D_{1,2}f^+_{0.1}-{1\over 2}f^+_{0.1}D_{1,2}f^-_{1.0}\over f^-_{1.0}}-f^+_{1.1}f^-_{1.0}+
{2(f^-_{1.1})^2f^+_{1.3}+(f^+_{0.1})^2f^-_{1.1}+f^-_{2.3}f^+_{0.1}f^+_{1.3}\over 2
f^-_{1.0}}
$$ 
$$
\tilde f^+_{1.3}={f^-_{1.0}D_{1,2}f^+_{1.3}-{1\over 2}f^+_{1.3}D_{1,2}f^-_{1.0})\over f^-_{1.0}}+f^+_{2.3}f^-_{1.0}+{(f^+_{1.3})^2f^-_{2.3}-3f^+_{1.3}f^-_{1.1}f^+_{0.1}-(f^+_{0.1})^3\over 2f^-_{1.0}}
$$
$$
\tilde f^-_{1.1}={f^-_{1.0}D_{1,2}f^-_{1.1}-{1\over 2}f^-_{1.1}D_{1,2}f^-_{1.0}\over f^-_{1.0}}+f^-_{0.1}f^-_{1.0}-{2(f^+_{0.1})^2f^-_{2.3}+(f^-_{1.1})^2f^+_{0.1}+f^-_{2.3}f^-_{1.1}f^+_{1.3}\over 2f^-_{1.0}}
$$ 
$$
{\tilde f^-_{1.0}\over f^-_{1.0}}={1\over 4}D_{1,2}^2 \ln f^-_{1.0}+f^-_{1.0}f^+_{1.0}+{1\over 2}(f^-_{0.1}f^+_{0.1}+f^-_{1.1}f^+_{1.1})+{3\over 2}(f^-_{2.3}f^+_{2.3}+f^-_{1.3}f^+_{1.3})+
$$
$$
{3\over 4}{f^+_{0.1}D_{1,2}f^-_{1.1}-f^-_{1.1}D_{1,2}f^+_{0.1}\over f^-_{1.0}}-{1\over 
4}{f^+_{1.3}D_{1,2}f^-_{2.3}-f^-_{2.3}D_{1,2}f^+_{1.3}\over f^-_{1.0}}-{f^+_{0.1}\over f^-_{1.0}}(f^-_{2.3}f^+_{1.2}-f^+_{0.1}f^-_{1.2})-
$$
$$
{1\over 4}{3(f^-_{1.0}f^+_{0.1})^2-(f^-_{2.3}f^+_{1.3})^2+6f^-_{1.0}f^+_{0.1})f^-_{2.3}f^+_{1.3}+4(f^-_{1.1})^3f^+_{1.3}+4(f^+_{0.1})^3f^-_{2.3}\over (f^-_{1.0})^2}
$$

\subsection{Discrete transformation of the complicate root $T_{\alpha_1+3\alpha_2}$}

The calculations below are not necessary if one pay attention on symmetry of the main system of equations (\ref{SG}) with respect to the following exchange of variables and unknown functions 
$$
D_{2,3}\to -D_{2,3},\quad D_{1.3}\to -D_{1,0},\quad D_{1,2}\to -D_{1,1},\quad D_
{0,1}\to D_{0,1},
$$
$$
f^{\pm}_{2.3}\to -f^{\pm}_{2.3},\quad f^{\pm}_{1.3}\to -f^{\pm}_{1.0},\quad
f^{\pm}_{1.1}\to -f^{\pm}_{1.2},\quad f^{\pm}_{0.1}\to -f^{\mp}_{0.1}
$$
Using this substitution or by straitforward calculations for $T_{\alpha_1+3\alpha_2}$ we obtain
$$
\tilde f^+_{1.3}={1\over f^-_{1.3}},\quad \tilde f^+_{0.1}={f^-_{1.2}\over f^-_{1.3}},\quad \tilde f^+_{1.2}=-{f^-_{0.1}\over f^-_{1.3}},\quad \tilde f^-_{1.0}={f^-_{2.3}\over f^-_{1.3}},\quad \tilde f^+_{2.3}=-{f^+_{1.0}\over f^-_{1.3}}
$$
$$
\tilde f^+_{1.1}=f^+_{1.1}+{(f^-_{0.1})^2+f^-_{1.2}f^+_{1.0}\over f^-_{1.3}},\quad
\tilde f^-_{1.1}=f^-_{1.1}+{-
(f^-_{1.2})^2+f^-_{2.3}f^-_{0.1}\over f^-_{1.3}}
$$
$$
\tilde f^-_{2.3}={-f^-_{1.3}D_{1,1}f^-_{2.3}+{1\over 2}f^-_{2.3}D_{1,1}f^-_{1.3}\over f^-_{1.3}}+f^-_{1.0}f^-_{1.3}+{(f^-_{2.3})^2f^+_{1.0}-(f^-_{1.2})^3+3f^-_{2.3}f^-_{1.1}f^-_{1.2}\over 2f^-_{1.3}}
$$
$$
\tilde f^+_{1.0}={-f^-_{1.3}D_{1,1}f^+_{1.0}+{1\over 2}f^+_{1.0}D_{1,1}f^-_{1.3}\over f^-_{1.3}}-f^+_{2.3}f^-_{1.3}-{f^-_{2.3}(f^+_{1.0})^2+2(f^-_{0.1})^3+3f^-_{0.1}f^-_{1.2}f^+_{1.2}\over 2f^-_{1.3}}
$$
$$
\tilde f^-_{1.2}={-f^-_{1.3}D_{1,1}f^-_{1.2}+{1\over 2}f^-_{1.2}D_{1,1}f^-_{1.3}\over f^-_{1.3}}-f^+_{0.1}f^-_{1.3}+{f^-_{2.3}(f^-_{0.1})^2+2(f^-_{0.1})^2f^-_{2.3}-f^-_{0.1}(f^-_{1.2})^2\over 2f^-_{1.3}}
$$
$$
\tilde f^-_{0.1}={-f^-_{1.3}D_{1,1}f^-_{0.1}+{1\over 2}f^-_{0.1}D_{1,1}f^-_{1.3}\over f^-_{1.3}}-f^+_{1.2}f^-_{1.3}+{f^-_{1.2}(f^-_{0.1})^2+2(f^-_{1.2})^2f^+_{1.0}-f^-_{0.1}f^-_{2.3}f^+_{1.0})\over 2f^-_{1.3}}
$$
$$
{\tilde f^-_{1.3}\over f^-_{1.3}}={1\over 4}D_{1,1}^2 \ln f^-_{1.3}+f^-_{1.1}f^+_{1.3}+{1\over 2}(f^-_{0.1}f^+_{0.1}+f^-_{1.2}f^+_{1.2})+{3\over 2}(f^-_{2.3}f^+_{2.3}+f^-_{1.0}f^+_{1.0})-
$$
$$
{3\over 4}{f^-_{0.1}D_{1,1}f^-_{1.2}-f^-_{1.2}D_{1,1}f^-_{0.1}\over f^-_{1.3}}-{1\over 4}{f^+_{1.0}D_{1,1}f^-_{2.3}-f^-_{2.3}D_{1,1}f^+_{1.0}\over f^-_{1.3}}+{f^-_{0.1}\over f^-_{1.3}}(f^-_{2.3}f^+_{1.1}+f^-_{0.1}f^-_{1.1})-
$$
$$
{1\over 4}{3(f^-_{1.3}f^-_{0.1})^2-(f^-_{2.3}f^+_{1.0})^2-6f^-_{1.3}f^-_{0.1})f^-_{2.3}f^+_{1.0}+4(f^-_{1.2})^3f^+_{1.3}-4(f^-_{0.1})^3f^-_{2.3}\over (f^-_{1.3})^2}
$$
Of course all formulae above for $T_1$ and $T_{\alpha_1+3\alpha_2}$ coinsides with the general once in the case of arbitrary semisimple algebra of the paper \cite{I}.

\subsection{General formulae of solution}

In spite of comlicate on the first look structure of dicrete transformations of the previous two subsections their resolving is observable. We would not like in this paper give the proofs but present only finally result of not simple calculations. Aritrary discrete transformation can be represented as consequent applications of two mutualy commutative basis transformations $T_1$ and $T_{3\alpha_2}\equiv T^{-1}_1T_{\alpha_1+3\alpha_2}$ and has the form $T=T^{n_1}_1T_{3\alpha_2}^{n_2}$. 
The result of calculation of discretre transformation $T_{3\alpha_2}^{n_2}T_1^{n_1}$ may be expessed in terms of the basis function (all factorial factors are included in definition of Wandermond determinants)
$$
V(n_1;n_2,n_3,n_4)= \int \prod_{k=1}^{n_1}P(\lambda_k) d\lambda_k \prod_{i=1}^{n_2}Q(\mu_i) d\mu_i\prod_{j=1}^{n_3}Q(\nu_j) d\nu_j \prod_{l=1}^{n_4}Q(\sigma_l) d\sigma_l \times
$$
$$
{W^2(\lambda)_{n_1}W^2(\mu)_{n_2}W^2(\nu)_{n_3}W^2(\sigma)_{n_4}\over \prod_{k=1}^{n_1}\prod_{i=1}^{n_2}\prod_{j=1}^{n_3}\prod_{l=1}^{n_4}(\lambda_k-\mu_i)(\lambda_k-\nu_j)(\lambda_k-\sigma_l)}
$$
In this notations solution of 6-wave $G_2$ problem looks similar as it was found before for the case of $A_2$ and $B_2$ algebras
$$
f^+_{1.0}={V(n_1-1;n_2,n_2,n_2)\over V(n_1;n_2,n_2,n_2)},\quad f^+_{0.1}={V(n_1;n_2-1,n_2,n_2)\over V(n_1;n_2,n_2,n_2)}
$$
$$
f^+_{1.1}={V(n_1-1;n_2-1,n_2,n_2)\over V(n_1;n_2,n_2,n_2)},\quad f^+_{1.2}={V(n_1-1;n_2-1,n_2-1,n_2)\over V(n_1;n_2,n_2,n_2)}
$$
$$
f^+_{1.3}={V(n_1-1;n_2-1,n_2-1,n_2-1)\over V(n_1;n_2,n_2,n_2)},\quad f^+_{2.3}={V(n_1-2;n_2-1,n_2-1,n_2-1)\over V(n_1;n_2,n_2,n_2)}
$$
\begin{equation}
f^-_{1.0}={V(n_1+1;n_2,n_2,n_2)\over V(n_1;n_2,n_2,n_2)},\quad f^-_{0.1}={V(n_1;n_2+1,n_2,n_2)\over V(n_1;n_2,n_2,n_2)}\label{W}
\end{equation}
$$
f^-_{1.1}={V(n_1+1;n_2+1,n_2,n_2)\over V(n_1;n_2,n_2,n_2)},\quad f^-_{1.2}={V(n_1+1;n_2+1,n_2+1,n_2)\over V(n_1;n_2,n_2,n_2)}
$$
$$
f^-_{1.2}={V(n_1+1;n_2+1,n_2+1,n_2)\over V(n_1;n_2,n_2,n_2)},\quad f^-_{1.3}={V(n_1+1;n_2+1,n_2+1,n_2+1)\over V(n_1;n_2,n_2,n_2)}
$$
$$
f^-_{2.3}={V(n_1+2;n_2+1,n_2+1,n_2+1)\over V(n_1;n_2,n_2,n_2)}
$$
Of course the formulae above on the level of the present paper it isnecessary consider as hipothesis and possible happy guess for solution of problem in the case of arbitrary semisimple algebra.

\section{Multisoliton solutions}

The explicit form of solution in the case of $B_2$ and $G_2$ allow to find conditions of the chain interupting and construct multisoliton solutions as it was done in the case of $A_2$ algebra \cite{TOK},\cite{TOK1} 

\section{Outlook}

The main result of the present paper consits in assuarence that calculations of this kind may done in the case of arbitrary semisimple algebra. It is possible to assume that to this aim is sufficient to resolve the discrete transformation chains of the simple roots of the algebra.
The finally result will be obtained by simple multiplications with taking into account only differences of spectral parameters of simple roots connected on the sheam of Dynkin.

The most intrigued is the fact that soliton solutions of multidimensional integrable system leads to the same limitations on paramaters amplituda-phase (with correction on interaction) as in the case $A_1$ algebra.  

The same surprising is existence of universal integrable chains in the field of Toda lattice.
Of course there description and understanding of their nature is very interesting problem
stated by the present paper. 

Ussualy all results in the theory of integrable system may be interpreted as some equalities form the theory of representtion of semisimple algebras (relations between highest vector of different irreducible representations and so on). What kind problems of representation theory  (if any) may be connected with arised equalities of the present paper?

At last the parameters $\lambda,\mu,\nu_i$ (the last after taking into account homogineous part
solutions (section 2) of all elements of lower triangular algebra) in comparisen with $A_1$ algebra case play role of spectral parameters. By this reason it is possible to assume that one-dimentional Lax formalism it is necessary to change on multidimensional one. This assumtion it is possible to put into connection with the fact that discrete transformation are always commutative and number of basis transformation coincides with the rank of the algebra.

All this problems demand for its solution further investigation. 

\section{Appendix}

In this Appendix we present detail calculations for the most complicate case of discrete transformation $T_{2\alpha_2}$ of $B_2$ algebra. We change greek indexes on latina ones here.

\subsection{Initial condition. Zero order case}

$$
A^0=\int {dm M dl L\over l-m},\quad B^0=\int {dm_1 M_1 dm_2 M_2 dl L\over (l-m_1)(l-m_2)}
$$

\subsection{First step of $T_{2\alpha_2}$ transformation}

$$
A^1={{1\over 2}D_1B^0_1-B^0(D_1)_1\over D_0}=\int {dm_1 M_1 dm_2 M_2 dm_3 M_3 dl L({m_1+m_2\over 2}-m_3)\over (l-m_1)(l-m_2)}=
$$
$$
\int {dm_1 M_1 dm_2 M_2 dm_3 M_3 dl L(m_1-m_3)\over (l-m_1)(l-m_2)}=
$$
$$
{1\over 2}\int dm_1 M_1 dm_2 M_2 dm_3 M_3 dl L{(m_1-m_3)\over(l-m_2)}({1\over l-m_1}-{1\over l-m_3})=
$$
$$
{1\over 2}\int dm_1 M_1 dm_2 M_2 dm_3 M_3 dl L{(m_1-m_3)^2\over(l-m_2)(l-m_1)(l-m_3)}
$$
In transformation above it was used only the fact the symmetry of the domain of integration with respect to permutation of all indexes $m_1,m_2,m_3$. 
Under calculation of $B^1$ from (\ref{CB}) we will not conserve indexes of integrale and differentiales remaining only integrant function. We have consequently
$$
{1\over 4}{(m_1+m_2)^2\over (l-m_1)(l-m_2)}+{m_4(m_3-(m_1+m_2))\over (l-m_1)(l-m_2)}-
{1\over 2}{(m_1-m_2)\over (l-m_1)}=
$$
$$
{m_1(m_2-m_4)+m_4(m_3-m_2)\over (l-m_1)(l-m_2)}={1\over 2}[{m_1(m_2-m_4)^2\over (l-m_1)(l-m_2)(l-m_4)}-{m_4(m_2-m_3)^2\over (l-m_1)(l-m_2)(l-m_3)}=
$$
$$
{1\over 2}{(m_1-m_4)(m_2-m_3)^2\over (l-m_1)(l-m_2)(l-m_3)}={1\over 4}{(m_1-m_4)^2(m_2-m_3)^2\over (l-m_1)(l-m_2)(l-m_3)(l-m_4)}
$$
And thus for $B^1$ we obtain
$$
B^1={1\over 4}\int dm_1 M_1 dm_2 M_2 dm_3 M_3 dm_4 M_4 dl L{(m_1-m_4)^2(m_2-m_3)^2\over(l-m_2)(l-m_1)(l-m_3)(l-m_4)}
$$
In the process of calculations above we have used only the fact symmetry of region of integration with respect to permutations of four parameters $m_i$
$$
A^2={1\over 8}{(m_5-m_6)^2(m_1-m_4)^2(m_2-m_3)^2\over (l-m_1)(l-m_2)(l-m_3)(l-m_4)}[{1\over 2}
(m_1+m_2+m_3+m_4)-(m_5+m_6)]=
$$
$$
{1\over 4}{(m_5-m_6)^2(m_1-m_4)^2(m_2-m_3)^2\over (l-m_1)(l-m_2)(l-m_3)(l-m_4)}[m_1-m_5]
$$
Now it is necessary to use equality
$$
(m_i-m_k)(m_j-m_l)=(m_j-m_k)(m_i-m_l)-(m_j-m_i)(m_k-m_l)
$$
with help of which we come to the equality 
$$
{1\over 4}{(m_5-m_6)^2(m_1-m_4)^2(m_2-m_3)^2\over (l-m_1)(l-m_2)(l-m_3)(l-m_4)}[m_1-m_5]=
$$
$$
{1\over 2}{(m_2-m_3)^2(m_1-m_4)(m_5-m_6)(m_1-m_6)(m_5-m_4)(m_1-m_5)\over (l-m_1)(l-m_2)(l-m_3)(l-m_4)}=
$$
$$
{1\over 2}{(m_2-m_3)^2(m_5-m_6)(m_1-m_6)(l-m_5)W_3(m_1,m_4,m_5)\over (l-m_1)(l-m_2)(l-m_3)(l-m_4)(l-m_5)}=
$$
$$
{1\over 2!3!}{(m_2-m_3)^2W^2_3(m_1,m_4,m_5)\over (l-m_1)(l-m_2)(l-m_3)(l-m_4)(l-m_5)}
$$
By induction we obtain
$$
A^n={1\over n!(n+1)!} \int Ldl \prod_{i=1}^{(2n+1)}M_i dm_i {W^2_n(m_1,..m_n)W^2_{n+1}(m_{n+1},..m_{2n+1})\over \prod_{i=1}^{(2n+1)}(l-m_i)}
$$
$$
B^n={1\over (n+1)!(n+1)!}\int Ldl \prod_{i=1}^{(2n+2)}M_i dm_i {W^2_{n+1}(m_1,..m_{n+1})W^2_{n+1}(m_{n+2},..m_{2n+2})\over \prod_{i=1}^{(2n+2)}(l-m_i)}
$$

\end{document}